\documentstyle[preprint,aps,psfig]{revtex}

\newcommand{\beqn}{\begin{eqnarray}}
\newcommand{\eeqn}{\end{eqnarray}}

\tightenlines

\begin{document}
\title{First Principles Justification of a ``Single Wave Model'' \\
  for Electrostatic Instabilities} \author{John David Crawford \and
  Anandhan Jayaraman}
\address{Department of Physics and Astronomy\\
  University of Pittsburgh\\
  Pittsburgh, Pennsylvania 15260} \date{\today} \maketitle

\begin{abstract}
  The nonlinear evolution of a unstable electrostatic wave is
  considered for a multi-species Vlasov plasma. From the singularity
  structure of the associated amplitude expansions, the asymptotic
  features of the electric field and distribution functions are
  determined in the limit of weak instability, i.e.
  $\gamma\rightarrow 0^+$ where $\gamma$ is the linear growth rate.
  The asymptotic electric field is monochromatic at the wavelength of
  the linear mode with a nonlinear time dependence. The structure of
  the distibutions outside the resonant region is given by the linear
  eigenfunction but in the resonant region the distribution is
  nonlinear.  The details depend on whether the ions are fixed or
  mobile; in either case the physical picture corresponds to the
  single wave model originally proposed by O"Neil, Winfrey, and
  Malmberg for the interaction of a cold weak beam with a plasma of
  fixed ions.

\end{abstract}
\vspace{0.5in}

\pacs{52.25.Dg, 47.20.Ky, 52.35.Fp, 52.35.Sb, 52.35.Qz}

\section{Introduction}

Recently, we have studied the collisionless nonlinear evolution of an
unstable mode; first in a single component Vlasov plasma with fixed
ions and then more generally in a multi-species Vlasov
plasma.\cite{jdc95}-\cite{jdcaj97} The asymptotic features of the
problem in the limit of weak instability, i.e. $\gamma\rightarrow0^+$
where $\gamma$ is the linear growth rate, were our principal focus.
The main tool has been expansions for the amplitude equation and the
distribution functions; in particular the asymptotic structure of
these expansions. Coefficients of both expansions develop
singularities as $\gamma\rightarrow0^+$ and these singularities reveal
the asymptotic features of the amplitude equation, distribution
functions and electric field.

The amplitude equation describes the evolution on the unstable
manifold of the equilibrium and a key conclusion of our previous paper
established the scaling behavior of this system.\cite{jdcaj97} By
setting $A(t)=\gamma^\beta r(\gamma t)\exp(-i\theta(t))$, the
resulting equations for $r(\tau)$ ($\tau\equiv\gamma t$) and
$\theta(t)$ were free of singular behavior as $\gamma\rightarrow0^+$,
provided the exponent $\beta$ was suitably chosen.  The correct choice
turned out to depend on the model under consideration: if ion masses
were finite, then typically $\beta=5/2$ unless the ion distributions
happened to be flat at the phase velocity of the linear wave. In the
limit of fixed ions ($m_i\rightarrow\infty$) or when the ion
distributions are flat at the resonant velocity, the exponent drops to
$\beta=1/2$.

In this paper we apply these results for $\beta$ to control and
interpret the singularities that arise in the expansions of the
distribution functions. This study illuminates in detail the
asymptotic structure of both the distributions and the electric field.
In particular, we find that the electric field is essentially
monochromatic at the wavelength of the linear mode with a nonlinear
time dependence. Outside the resonant region, the distributions are
described by the linear eigenfunction and in the resonant region they
have a nonlinear structure.  The details depend on whether the ions
are fixed or mobile, but in either case this physical picture is well
known from the ``single wave model'' proposed by O'Neil, Winfrey and
Malmberg for the interaction of cold weak beam with a neutral plasma
of mobile electrons and infinitely massive ions..\cite{owm} Their work
supplied a model of the self-consistent Vlasov problem that has proven
useful to many researchers in the subsequent
years.\cite{ow}-\cite{doveil} Our conclusions generalize this useful
simplified picture to a general electrostatic instability arising in
an unmagnetized multi-species Vlasov plasma.  As this paper was being
completed, we learned of the interesting recent work by
del-Castillo-Negrete who has given an different derivation of the
single wave picture using matched asymptotic methods to treat the
resonant and non-resonant particles.\cite{dcn} As in the original work
of O'Neil {\em et al.}, del-Castillo-Negrete allows only mobile
electrons and moreover restricts attention to instabilities associated
with so-called ``inflection point modes''.\cite{shad,jdc95b}

In the remainder of this introduction we review our notation and in
section \ref{sec:prior} we summarize the needed conclusions of Ref
\cite{jdcaj97} regarding the singularities of the expansions. The
third section applies these conclusions to the distributions and
electric field, and section \ref{sec:disc} contains a final
discussion.

\subsection{Notation}

Our notation follows Ref \cite{jdcaj97}; we consider a 
one-dimensional, multi-species Vlasov plasma defined by
\begin{equation}
  \frac{\partial F^{(s)}}{\partial t}+v\frac{\partial
    F^{(s)}}{\partial x}+ \kappa^{(s)}\;E\;\frac{\partial
    F^{(s)}}{\partial v}=0\label{eq:vlasov}
\end{equation}
\begin{equation}
  \frac{\partial E}{\partial
    x}=\sum_s\int^\infty_{-\infty}\,dv\,F^{(s)}(x,v,t).\label{eq:poisson}
\end{equation}
Here $x$, $t$ and $v$ are measured in units of $u/\omega_e$,
$\omega_e^{-1}$ and $u$, respectively, where $u$ is a chosen velocity
scale and $\omega_e^2=4\pi e^2n_e/m_e$.  The plasma length is $L$ with
periodic boundary conditions and we adopt the normalization
\begin{equation}
  \int^{L/2}_{-L/2}\,dx\,\int^\infty_{-\infty}\,dv\,F^{(s)}(x,v,t)=
  \left(\frac{z_s\;n_s}{n_e}\right)L\label{eq:Fnorm}
\end{equation}
where $q_s=e\,z_s$ is the charge of species $s$ and
$\kappa^{(s)}\equiv {q_sm_e/em_s}$. Note that $\kappa^{(e)}=-1$ for
electrons and that the normalization (\ref{eq:Fnorm}) for negative
species makes the distribution function negative.

Let $F_0(v)$ and $f(x,v,t)$ denote the multi-component fields for the
equilibrium and perturbation respectively and $\kappa$ the matrix of
mass ratios,
\begin{equation}
  f\equiv\left(\begin{array}{c} f^{(s_1)}\\f^{(s_2)}\\
      \vdots\end{array}\right)\;\;\;\;
  F_0\equiv\left(\begin{array}{c}F_0^{(s_1)}\\F_0^{(s_2)}\\
      \vdots\end{array}\right)\;\;\;\;
  \kappa\equiv\left(\begin{array}{cccc}\kappa^{(s_1)} & 0 & 0 & \cdots\\
      0&\kappa^{(s_2)}&0&\cdots\\
      \vdots&\vdots&\vdots\end{array}\right),
\end{equation}
then the system (\ref{eq:vlasov}) - (\ref{eq:poisson}) can be
concisely expressed as
\begin{equation}
  \frac{\partial f}{\partial
    t}={\cal{L}}\,f+{\cal{N}}(f)\label{eq:model}
\end{equation}
where the linear operator is defined by
\begin{eqnarray}
  {\cal{L}}\,f&=&\sum^{\infty}_{l=-\infty}\,e^{ilx}\,(L_l f_l)(v)
\label{eq:fexp}\\
(L_l f_l)(v)&=&\left\{\begin{array}{cc}0&l=0\\
    -il\left[vf_l(v)+\kappa\cdot\eta_l(v)\sum_{s}
      \int^\infty_{-\infty}\,dv'\,f^{(s)}_l(v')
    \right]&l\neq0,\end{array}\right.\label{eq:linop}
\end{eqnarray}
with $\eta_l(v)\equiv-\partial_vF_0/l^2$, and the nonlinear operator
${\cal{N}}$ is
\begin{equation}
  {\cal{N}}(f)=\sum^{\infty}_{m=-\infty}\,
  e^{imx}\,{\sum^{\infty}_{l=-\infty}}'\, \frac{i}{l}\left(\kappa\cdot
    \frac{\partial f_{m-l}}{\partial v}\right)
  \sum_{s'}\int^\infty_{-\infty}\,dv'\,f^{(s')}_l(v').
\label{eq:nop}
\end{equation}
In the spatial Fourier expansion (\ref{eq:fexp}), $l$ denotes an
integer multiple of the basic wavenumber $2\pi/L$, and the primed
summation in (\ref{eq:nop}) omits the $l=0$ term. The notation
$\kappa\cdot\eta_l(v)$ or $\kappa\cdot\partial_v f_{m-l}$ denotes
matrix multiplication.
For two multi-component fields of $(x,v)$, e.g.
$B=(B^{(s_1)},B^{(s_2)},B^{(s_3)},..)$ and
$D=(D^{(s_1)},D^{(s_2)},D^{(s_3)},..)$, we define an inner product
by
\begin{eqnarray}
  (B,D)&\equiv&\sum_s\int^{L/2}_{-L/2}\,dx\int_{-\infty}^{\infty}\,dv\,
  B^{(s)}(x,v)^\ast D^{(s)}(x,v) =\int^{L/2}_{-L/2}\,dx\,<B,D>
\end{eqnarray}
where
\begin{eqnarray}
  <B,D>&\equiv&\sum_s \int_{-\infty}^{\infty}\,dv\,B^{(s)}(x,v)^\ast
  D^{(s)}(x,v).
\end{eqnarray}

The spectral theory for ${\cal{L}}$ is well established and the
facts needed for our analysis are easily summarized. 
The eigenvalues $\lambda=-il z$ of
${\cal{L}}$ are determined by the roots $\Lambda_{l}(z)=0$ of the
``spectral function'',
\begin{equation}
  \Lambda_{l}(z)\equiv 1+
  \int^\infty_{-\infty}\,dv\,\frac{\sum_s\kappa^{(s)}\eta_l^{(s)}(v)}{v-z}.
\label{eq:specfcn}
\end{equation}
If the contour in (\ref{eq:specfcn}) is replaced by the Landau contour
for ${\rm Im}(z)<0$ then we have the linear dielectric
$\epsilon_{{l}}(z)$; for ${\rm Im}(z)>0$, $\Lambda_{l}(z)$ and
$\epsilon_{{l}}(z)$ are the same function.  The eigenvalues can
be either real or complex depending on the symmetry and shape of the
equilibrium.

Associated with an eigenvalue $\lambda=-il z$ is the multi-component
eigenfunction $\Psi(x,v)=e^{ilx}\,\psi(v)$ where
\begin{equation}
\psi(v)=-\frac{\kappa\cdot \eta_l}{v-z}.
\end{equation}
There is also an associated adjoint eigenfunction
$\tilde{\Psi}(x,v)=e^{ilx}\tilde{\psi}(v)/L$ satisfying
$(\tilde{\Psi},\Psi)=1$ with
\begin{equation}
  \tilde{\psi}(v)=- \frac{1}{\Lambda'_{l}(z)^\ast(v-z^\ast)}.
\label{eq:aefcn2}
\end{equation}
Note that all components of $\tilde{\psi}(v)$ are the same.  The
normalization in (\ref{eq:aefcn2}) assumes that the root of
$\Lambda_{l}(z)$ is simple and is chosen so that
$<\tilde{\psi},\psi>=1$.  The adjoint determines the projection of
$f(x,v,t)$ onto the eigenvector, and this projection defines the
time-dependent amplitude of $\Psi$, i.e.
$A(t)\equiv(\tilde{\Psi},f)$.

\section{Previous results }\label{sec:prior}

The equilibrium $F_0(v)$ is assumed to support a ``single'' unstable
mode in the sense that $E^u$, the unstable subspace for ${\cal L}$, is
two-dimensional. With translation symmetry and periodic boundary
conditions, this is the simplest instability problem that can be
posed.  Henceforth, let $k$ denote the wavenumber of this unstable
mode that is associated with the root $\Lambda_{k}(z_0)=0$ which we
assume to be simple, i.e. $\Lambda'_{k}(z_0)\neq0.$ The corresponding
eigenvector is
\begin{equation}
  \Psi(x,v)=e^{ikx}\,\psi(v)= e^{ikx}\left(-\frac{\kappa\cdot
      \eta_k}{v-z_0}\right).
\label{eq:lefcn}
\end{equation}
The root $z_0=v_p+{i\gamma}/{k}$ determines the phase velocity
$v_p=\omega/k$ and the growth rate $\gamma$ of the linear mode as the
real and imaginary parts of the eigenvalue
$\lambda=-ikz_0=\gamma-i\omega$.

Solutions on the unstable manifold have the form
\begin{eqnarray}
  f^u(x,v,t) = \left[ A(t) \psi(v) e^{ikx} + A^*(t) \psi^*(v)
    e^{-ikx} \right] + H(x,v,A(t),A^*(t))\label{eq:funstab}
\end{eqnarray}
where $A(t)\equiv(\tilde{\Psi},f^u)$ evolves according to the
amplitude equation
\begin{equation}
\dot{A}=\lambda\,A+(\tilde{\Psi},{\cal{N}}(f^u))\label{eq:adot}
\end{equation}
and self-consistency requires $H$ to satisfy
\begin{eqnarray}
 \frac{\partial
    H}{\partial A} \dot A + \frac{\partial H}{\partial A^*} \dot {A^*}
  &=&{\cal{L}}
  H+{\cal{N}}(f^u)-\left[(\tilde{\Psi},{\cal{N}}(f^u))\,\Psi +
    cc\right]\label{eq:Sdotwu}
\end{eqnarray}
subject to the geometric constraints
\begin{equation}
  0=H(x,v,0,0)=\frac{\partial H}{\partial A}(x,v,0,0)= \frac{\partial
    H}{\partial A^\ast}(x,v,0,0).\label{eq:tang}
\end{equation}

The translation symmetry of the model (\ref{eq:model}) provides
important constraints on both the amplitude equation and the form of
$H$.\cite{jdcaj97} For the amplitude equation (\ref{eq:adot}), the
right hand side must have the form
\begin{eqnarray}
\lambda A + (\tilde{\Psi},{\cal{N}}(f^u)) = A p(\sigma)\label{eq:pdef}
\end{eqnarray}
where $\sigma\equiv |A|^2$ and $p(\sigma)$ is an unknown function to
be determined from the model. Similarly, translational symmetry
requires the spatial Fourier components of $H$ to have a special form
\begin{eqnarray}
  H_0(v,A,A^\ast)&=&\sigma\,h_0(v,\sigma)\nonumber\\
  H_k(v,A,A^\ast)&=&A\sigma\,h_1(v,\sigma)\label{eq:hdef}\\
  H_{mk}(v,A,A^\ast)&=&A^m\,h_m(v,\sigma)\;\;\;\;
  {\mbox{for}}\;\;m\geq2\nonumber
\end{eqnarray}
where $H_{-l}={H_l}^\ast$. These results focus our analysis on a set
of functions, $\{p(\sigma), h_m(v,\sigma)\}$, which must be determined
from the Vlasov equation.

\subsection{Expansions and singularities}
  
We study $p(\sigma)$ and $\{h_m(v,\sigma)\}$ via the expansions
\begin{equation}
  p(\sigma) = \sum_{j=1}^{\infty} p_j \sigma^j  \hspace{0.5in}
  h_m(v,\sigma) = \sum_{j=1}^{\infty} h_{m,j}(v) \sigma^j.\label{eq:pjdef}
\end{equation}
The coefficients $p_j$ and $h_{m,j}$ are determined by inserting the
expansions into (\ref{eq:Sdotwu}) and (\ref{eq:pdef}) 
and solving at each order of $\sigma$. The resulting recursion
relations are given in Ref \cite{jdcaj97} and are not required 
for the present discussion. 

The key point is that for both the
amplitude equation and the distribution function the expansion
coefficients develop singularities in the limit $\gamma\rightarrow0^+$.
This can be seen explicitly by reviewing the calculation of the 
cubic coefficient $p_1$. From Ref \cite{jdcaj97}, $p_1$ depends
on $h_{0,0}$ and $h_{2,0}$,
\begin{equation}
p_1=-\frac{i}{k}\left[<\partial_v\tilde{\psi},\kappa\cdot(h_{0,0}-h_{2,0})>
+\frac{\Gamma_{2,0}}{2}<\partial_v\tilde{\psi},\kappa\cdot\psi^\ast>\right],
\label{eq:p1coeff}
\end{equation}
where $\Gamma_{2,0}=\int dv\, h_{2,0}$. The recursion relations determine
$h_{0,0}$ and $h_{2,0}$,
\begin{eqnarray}   
h_{0,0}(v)&=& -\frac{1}{k^2}\frac{\partial}{\partial v}
\left[\frac{\kappa^2\cdot\eta_k}{(v-z_0)(v-z^\ast_0)} \right]
\label{eq:h00}\\
h_{2,0}(v)&=&\frac{1}{2k^2} \left(\frac{\kappa\cdot\partial_v
 \psi}{v-z_0}\right) +\frac{1}{6
k^2}\left(\frac{\kappa\cdot\eta_k}{v-z_0}\right) \left(\frac{
 \kappa\cdot\eta_k}{v-z_0}\right), \label{eq:h20}
\end{eqnarray}
and one notes that for $\gamma>0$ these are smooth functions but there
are complex poles at $z_0$ and $z_0^\ast$ that approach the real axis
at $v=v_p$ in the limit  $\gamma\rightarrow0^+$. For  $h_{2,0}$ all poles
lie above the real axis, but $h_{0,0}$ contains poles above and below the
axis and this forces the integral 
$<\partial_v\tilde{\psi},\kappa\cdot h_{0,0}>$ in $p_1$ to diverge as
$\gamma\rightarrow0^+$ because of a pinching singularity. For similar 
reasons,
the integral $<\partial_v\tilde{\psi},\kappa\cdot\psi^\ast>$ also diverges
but the remaining integrals in $p_1$ are nonsingular. 

A detailed evaluation
of this asymptotic structure in $p_1$ yields the form
\begin{eqnarray}
  p_1&=&\frac{1}{\gamma^4} \left[c_1(\gamma)-\gamma\,d_1(\gamma) +
    {\cal O}(\gamma^2)\right]
\label{eq:p1asymp}
\end{eqnarray}
where $c_1$ and $d_1$ are nonsingular functions of $\gamma$ defined by
\begin{eqnarray}
  c_1(\gamma)&=&-\frac{k}{4\Lambda_k'(z_0)}
  {\sum_s}'{\kappa^{(s)}(1-{\kappa^{(s)}}^2)}\;{\rm
    Im}\left(\int^\infty_{-\infty}\,dv\frac{\eta_k^{(s)}}{v-z_0}\right)
\label{eq:a1}\\
d_1(\gamma)&=&\frac{1}{4}-
\frac{1}{4\Lambda_k'(z_0)}{\sum_s}'\kappa^{(s)}(1-{\kappa^{(s)}}^2)
\int^\infty_{-\infty}\,dv\frac{\eta_k^{(s)}}{(v-z_0)^2}\label{eq:b1}
\end{eqnarray}
where the primed species sum omits the electrons. 
At $\gamma=0$, $c_1$ has the limit
\begin{equation}
c_1(0)=-\frac{\pi k}{4\Lambda_k'(z_0)}
  {\sum_s}'{\kappa^{(s)}(1-{\kappa^{(s)}}^2)}\eta_k^{(s)}(v_p)
\end{equation}
which is typically non-zero yielding a $\gamma^{-4}$ singularity for
$p_1$. There are at least two special cases of interest for which
$c_1(0)=0$; namely, infinitely massive fixed ions 
($\kappa^{(s)}=0$ for all $s\neq e$) and flat ion distributions at the
resonant velocity ($\eta_k^{(s)}(v_p)=0$ for
all $s\neq e$). In such cases, the divergence of $p_1$ drops to 
$\gamma^{-3}$.

Analagous singularities appear also in the higher order coefficients and
grow more severe although their character remains the same. The higher
coefficients $h_{m,j}$ exhibit more and more poles which approach
the linear phase velocity as $\gamma\rightarrow0^+$ and these poles
generate stronger pinching singularities in the higher coefficients $p_j$.
An important property of the poles in $h_{m,j}$ is that they always 
have the general form $(v-\alpha)^{-n}$ or $(v-\alpha^\ast)^{-n}$ with
\begin{equation}
\alpha=z_0+i\gamma\zeta/k=v_p+i\gamma(\zeta+1)/k\label{eq:poles}
\end{equation}
where $\zeta>0$ is a purely numerical factor, 
i.e. the poles always lie along the vertical line ${\rm  Re}(v)=v_p$.

The explicit calculation of higher order coefficients from recursion
relations rapidly becomes prohibitively laborious; however, useful bounds
on the singularity of the higher order coefficients are obtained 
Ref \cite{jdcaj97} using an induction argument. More precisely, we find
for the amplitude equation
\begin{equation}
\lim_{\gamma\rightarrow0^+} \gamma^{\nu} |p_j| < \infty
\label{eq:pj}
\end{equation}
for $j\geq1$ where $\nu=5j-1$ in the generic case with 
$c_1(0)\neq0$, and $\nu=4j-1$
in the two special cases mentioned above, fixed ions or flat ion
distributions, with $c_1(0)=0$. 
For the coefficients of the distribution function, the
induction argument proves, for $m\geq0$, $j\geq0$ and $m' \ge 0$,
\begin{equation}
\lim_{\gamma\rightarrow0^+} \gamma^{\mu_{m,j}} \left|\int^\infty_{-\infty}
dv \, \sum_s \left(\kappa^{(s)}\right)^{m'} h^{(s)}_{m,j}(v)\right| < \infty
\label{eq:hmj}
\end{equation}
where $\mu_{m,j}=J_{m,j}+1$ with 
$J_{m,j}\equiv (2m + 5j -3) + 4\delta_{m,0} + 5\delta_{m,1}$
in the generic case defined by $c_1(0)\neq0$. For the special cases
of fixed ions or flat ion
distributions, (\ref{eq:hmj}) holds with exponent 
$\mu_{m,j}=J_{m,j}-j-\delta_{m,1}$.

The first bound determines the scaling exponent for $A$. 
When (\ref{eq:adot}), (\ref{eq:pdef}), and (\ref{eq:pjdef}) are combined 
we obtain an amplitude equation,
\begin{equation}
 \dot A =\lambda A + \sum_{j=1}^\infty p_j |A|^{2j}A
\label{eq:ampeq}
\end{equation}
where  each nonlinear term has a singular
coefficient and the equation is ill-defined as $\gamma\rightarrow0^+$. 
The cure is to rescale the amplitude 
\begin{equation}
A(t)\equiv \gamma^\beta\,r(\gamma t) e^{-i\theta(t)}\label{eq:rescale}
\end{equation}
with $\beta=5/2$ for the typical case ($c_1(0)\neq0$) and in the special
cases with $c_1(0)=0$ we require $\beta=2$.  Once this is done, the 
equations for $r(\tau)$ and $\theta(t)$ are nonsingular in the regime of weak 
growth rates;  additional details may be found in Ref \cite{jdcaj97}.

\section{Distribution functions and electric field}

The scaling (\ref{eq:rescale}) of the amplitude has immediate
implications for the asymptotic structure of the distributions. From
(\ref{eq:funstab}) the Fourier coefficients of $f=F-F_0$ may be
written in terms of $r(\tau)$ and $\theta(t)$ in (\ref{eq:rescale})
and $h_{m}$
\begin{eqnarray}
f_0(v,t)&=&\gamma^{2\beta}\,r(\tau)^2\,h_{0}(v,\gamma^{2\beta}r^2)\nonumber\\
f_k(v,t)&=&\gamma^{\beta}\,r(\tau)e^{-i\theta(t)}\;
\left[\psi(v)+\gamma^{2\beta}\,r(\tau)^2\,h_{1}(v,\gamma^{2\beta}r^2)\right]
\label{eq:fcomps}\\
f_{mk}(v,t)&=&\gamma^{2m\beta}\,r(\tau)^{m}\,e^{-im\theta(t)}\;
h_{m}(v,\gamma^{2\beta}r^2)\hspace{1.0in}m\geq2.\nonumber
\end{eqnarray}
As $\gamma\rightarrow0^+$, $r(\tau)$ is an ${\cal O}(1)$ quantity,
thus the asymptotic features of each Fourier component are determined
by the explicitly shown factors of $\gamma$ and the asymptotic form of
the functions $\psi(v)$ and $h_{m}(v,\gamma^{2\beta}r^2)$. The
dependence on $h_{m}$ necessitates a separate consideration of the
asymptotic behavior for non-resonant and resonant velocities. In the
former regime we assume the distance from the linear phase velocity
satisfies $v-v_p={\cal O}(1)$, and the resonant regime corresponds to
velocities within a neighborhood of $v_p$ that scales with the growth
rate, i.e. $v-v_p={\cal O}(\gamma)$.  For resonant velocities, the
singularities in $\psi(v)$ and $h_{m}(v,\gamma^{2\beta}r^2)$ come into
play and alter the asymptotic features of the distribution function.

\subsection{Non-resonant velocities}

For $v-v_p={\cal O}(1)$, the functions 
$\psi(v)$ and $h_{m}(v,\sigma)$ are bounded ${\cal O}(1)$ quantities (we use 
$\sigma=\gamma^{2\beta}r^2$ to emphasize this), and the Fourier components
(\ref{eq:fcomps}) combine to yield
\begin{eqnarray}
{\left(F(x,v,t)-F_0(v)\right)}/{\gamma^\beta}&=&
\left[r(\tau)e^{-i\theta(t)}\Psi(x,v)+\mbox{\rm cc}\right]+
\gamma^{\beta}\,r(\tau)^2\,h_{0}(v,\sigma)\\
&&+\gamma^{2\beta}r(\tau)^3\left[e^{-i\theta(t)}h_{1}(v,\sigma)e^{ikx}
+\mbox{\rm cc}\right]\nonumber\\
&&+\sum_{m=2}^{\infty}\left[\gamma^{(2m-1)\beta}e^{-im\theta(t)}\,
h_{m}(v,\sigma)e^{imkx}+\mbox{\rm cc}\right]\nonumber\\
&&\nonumber\\
&\approx& \left[r(\tau)e^{-i\theta(t)}\Psi(x,v)+\mbox{\rm cc}\right]+
{\cal O}(\gamma^{\beta}).
\label{eq:nonres}
\end{eqnarray}
In words, the non-resonant correction to $F_0$ scales overall as 
${\gamma^\beta}$; the leading piece of this correction simply has the form
of the linear wave $\Psi(x,v)$ with nonlinear time dependence determined
by the mode amplitude $r(\tau)\exp({-i\theta(t)})$.

\subsection{Resonant velocities}

For $v-v_p={\cal O}(\gamma)$, the functions 
$\psi(v)$ and $h_{m}(v,\sigma)$ typically develop singularities as 
$\gamma\rightarrow0^+$ and these divergences compete with the 
explicit factors of $\gamma$ in (\ref{eq:fcomps}) to determine the asymptotic
form of the distribution. The analysis is simplified by the fact
that all relevant singularities  are poles of the form described
in (\ref{eq:poles}), and these may be rewritten as a singular
factor multiplying a nonsingular function of the rescaled velocity
variable $u\equiv(v-v_p)/\gamma$, e.g.
\begin{equation}
\frac{1}{(v-\alpha)^n}=\frac{1}{\gamma^n}\frac{1}{((v-v_p)/\gamma-i(\zeta+1)/k)^n}
=\frac{1}{\gamma^n}\frac{1}{((u-i(\zeta+1)/k)^n}.
\end{equation}
Once this is done, the functions $\psi(v)$ and $h_{m}(v,\sigma)$, expressed
in terms of $u$, may be substituted into (\ref{eq:fcomps}); the 
variable $u$ provides a uniform velocity coordinate for the resonant region.

The puzzle is to deduce the correct overall factor of ${1}/{\gamma^n}$ 
for each function. For $h_{m}(v,\sigma)$ we have the integral 
bound (\ref{eq:hmj})
which may be rewritten in terms of $u$
\begin{equation}
\lim_{\gamma\rightarrow0^+}\left|\int^\infty_{-\infty}
du \, \sum_s \left(\kappa^{(s)}\right)^{m'} 
\gamma^{1+\mu_{m,j}} h_{m,j}(v_p+\gamma u)\right| < \infty.
\label{eq:hmju}
\end{equation}
Since all singularities are poles we know the integrand does not have an
integrable singularity, hence we conclude that 
$\gamma^{1+\mu_{m,j}} h_{m,j}(v_p+\gamma u)$ defines a nonsingular function
of $u$:
\begin{equation}
h_{m,j}(v_p+\gamma u)\equiv\gamma^{-(1+\mu_{m,j})}\hat{h}_{m,j}(u,\gamma).
\end{equation}
The nonsingular character of $\hat{h}_{m,j}(u,\gamma)$ can be checked
directly for the specific examples in (\ref{eq:h00}) and (\ref{eq:h20}), 
and also verified, in general, from the recursion relations. 
From the expansion of $h_{m}(v,\sigma)$, we thus find
\begin{eqnarray}
h_{m}(v_p+\gamma u,\sigma)&=&\sum_{j=0}^{\infty} \gamma^{2j\beta-(1+\mu_{m,j})}
\hat{h}_{m,j}(u,\gamma) r^{2j}.
\end{eqnarray}
In the generic case with $\beta=5/2$ and $\mu_{m,j}=J_{m,j}+1$ this
gives
\begin{eqnarray}
h_{m}(v_p+\gamma u,\sigma)&=&\frac{1}{\gamma^{\delta_m}}
\sum_{j=0}^{\infty} 
\hat{h}_{m,j}(u,\gamma) r^{2j}\label{eq:hmscale}
\end{eqnarray}
where
\begin{equation}
\delta_m=\left\{\begin{array}{lc}
3& m=0\\
6&m=1\\
2m-1&m\geq2
\end{array}\right.\hspace{0.5in}(c_1(0)\neq0).\label{eq:generic}
\end{equation}
In the special cases with fixed ions or flat distributions, then
$\beta=2$ and $\mu_{m,j}=J_{m,j}-j-\delta_{m,1}$, and
(\ref{eq:hmscale}) holds with exponent
\begin{equation}
\delta_m=\left\{\begin{array}{lc}
2& m=0\\
4&m=1\\
2m-2&m\geq2
\end{array}\right.\hspace{0.5in}(c_1(0)=0).\label{eq:special}
\end{equation}
In all cases, we define the nonsingular function 
$\hat{h}_{m}(u,r^2,\gamma)\equiv\sum_{j=0}^{\infty} 
\hat{h}_{m,j}(u,\gamma) r^{2j}$ and rewrite (\ref{eq:hmscale})
\begin{eqnarray}
h_{m}(v_p+\gamma u,\sigma)&=&\frac{\hat{h}_{m}(u,r^2,\gamma)}{\gamma^{\delta_m}}.
\label{eq:hmscal}
\end{eqnarray}

It is simpler to obtain the corresponding factorization of the eigenfunction;
from the definition (\ref{eq:lefcn}) we have
\begin{equation}
\psi(v_p+\gamma u)= \frac{1}{\gamma}\left(-\frac{\kappa\cdot
      \eta_k(v_p+\gamma u)}{u-i/k}\right),\label{eq:lefcnu}
\end{equation}
and the only subtlety concerns $\eta_k(v_p+\gamma u)$ which is ${\cal O}(1)$
in the generic case ($c_1(0)\neq0$) and 
${\cal O}(\gamma)$ in the two special cases with $c_1(0)=0$. Thus we define
the nonsingular function $\hat{\psi}(u,\gamma)$ by
\begin{equation}
\psi(v_p+\gamma u)= \frac{\hat{\psi}(u,\gamma)}{\gamma}
\hspace{0.5in}(c_1(0)\neq0),\label{eq:gefcn}
\end{equation}
in the generic case, but in the special cases the eigenfunction is
itself nonsingular and we have
\begin{equation}
\psi(v_p+\gamma u)= {\hat{\psi}(u,\gamma)}
\hspace{0.5in}(c_1(0)=0).\label{eq:scefcn}
\end{equation}
We are now able to describe the asymptotic structure of the distributions.

\subsubsection{Generic instability: $c_1(0)\neq0$}

For the generic case, inserting (\ref{eq:hmscal}) and (\ref{eq:gefcn})
into (\ref{eq:fcomps}) yields
\begin{eqnarray}
\lefteqn{{\left[F(x,v_p+\gamma u,t)-F_0(v_p+\gamma u)\right]}/{\gamma^{3/2}}=
\left\{r(\tau)\,e^{-i\theta(t)}\,e^{ikx}\left[\hat{\psi}(u,\gamma)
+r(\tau)^2\,\hat{h}_{1}(u,r^2,\gamma)\right]+\mbox{\rm cc}\right\}}\nonumber\\
&&\hspace{0.5in}+\sqrt{\gamma}\left\{r(\tau)^2\,\hat{h}_{0}(u,r^2,\gamma)
+\sum_{m=2}^{\infty}
\left[\gamma^{(m-2)/2}\,e^{imkx}\,r(\tau)^m\,
e^{-im\theta(t)}\,\hat{h}_{m}(u,r^2,\gamma)+\mbox{\rm cc}\right]
\right\};
\label{eq:resf}
\end{eqnarray}
neglecting the subdominant terms this gives
\begin{eqnarray}
\left[F(x,v_p+\gamma u,t)-F_0(v_p+\gamma u)\right]/\gamma^{3/2}&=&
\left\{r(\tau)\,e^{-i\theta(t)}\,e^{ikx}\left[\hat{\psi}(u,\gamma)
+r(\tau)^2\,\hat{h}_{1}(u,r^2,\gamma)\right]+\mbox{\rm cc}\right\}\nonumber\\
&&\hspace{1.0in}+{\cal O}(\sqrt{\gamma}).
\label{eq:resfd}
\end{eqnarray}
The generic resonant correction to $F_0$, expressed in the velocity
coordinate $u$, scales overall as $\gamma^{3/2}$; the leading term in
this correction has the wavelength of the linear wave but the velocity
dependence, $\hat{\psi}(u,\gamma)
+r(\tau)^2\,\hat{h}_{1}(u,r^2,\gamma)$, is not simply given by the
linear eigenfunction. The time dependence is determined by the mode
amplitude $r(\tau)\exp({-i\theta(t)})$ but the dependence on $r$ is
rather complicated.

\subsubsection{Special cases: $c_1(0)=0$}

For the special cases, defined by fixed ions or flat ion distributions,
we apply (\ref{eq:scefcn}) and  (\ref{eq:special})-(\ref{eq:hmscal})
to (\ref{eq:fcomps}) and obtain
\begin{eqnarray}
\lefteqn{{\left[F(x,v_p+\gamma u,t)-F_0(v_p+\gamma u)\right]}/{\gamma^{2}}=
\left\{r(\tau)\,e^{-i\theta(t)}\,e^{ikx}\left[\hat{\psi}(u,\gamma)
+r(\tau)^2\,\hat{h}_{1}(u,r^2,\gamma)\right]+\mbox{\rm cc}\right\}}\nonumber\\
&&\hspace{1.0in}+r(\tau)^2\,\hat{h}_{0}(u,r^2,\gamma)
+\sum_{m=2}^{\infty}
\left[e^{imkx}\,r(\tau)^m\,
e^{-im\theta(t)}\,\hat{h}_{m}(u,r^2,\gamma)+\mbox{\rm cc}\right].
\label{eq:nonresf}
\end{eqnarray}
This is a qualitatively different structure in contrast to (\ref{eq:resf});
now the resonant correction is ${\cal O}(\gamma^2)$ and {\em all}
wavelengths are present at leading order. Thus the spatial dependence
is very rich and bears no special relation to the linear instability; a 
similar observation holds for the dependence on velocity.

\subsection{Electric field}

The Fourier components of $E$ are given by (\ref{eq:fcomps}) and
Poisson's equation
\begin{eqnarray}
ikE_k(t)&=&\gamma^{\beta}\,r(\tau)e^{-i\theta(t)}\;
\left[1+\gamma^{2\beta}\,r(\tau)^2\,
\int_{-\infty}^\infty dv \sum_s h^{(s)}_{1}(v,\sigma)\right]
\nonumber\\
&&\label{eq:Ecomps}\\
imkE_{mk}(t)&=&\gamma^{2m\beta}\,r(\tau)^{m}\,e^{-im\theta(t)}\;
\int_{-\infty}^\infty dv \sum_s h^{(s)}_{m}(v,\sigma)
\hspace{1.0in}m\geq2.
\nonumber
\end{eqnarray}
Bounds on the asymptotic form of the integrals can be inferred from
the expansion $h_m=\sum_j h_{m,j}\sigma^j$ and the bound
(\ref{eq:hmj}) on the integrals of $h_{m,j}$ (for $m'=0$). The details
depend on whether we consider the generic instability or the special
cases.

\subsubsection{Generic instability: $c_1(0)\neq0$}

From (\ref{eq:hmj}) we find
\begin{equation}
\lim_{\gamma\rightarrow0^+} \gamma^{\alpha_m} \left|\int^\infty_{-\infty}
dv \, \sum_s h^{(s)}_{m}(v)\right| < \infty
\label{eq:hmg}
\end{equation}
with
\begin{equation}
\alpha_m=\left\{\begin{array}{lc}
2& m=0\\
5&m=1\\
2m-2&m\geq2
\end{array}\right.\hspace{0.5in}(c_1(0)\neq0).\label{eq:ageneric}
\end{equation}
Hence, with $\beta=5/2$, the generic components are
\begin{eqnarray}
ikE_k(t)&=&\gamma^{5/2}\,r(\tau)e^{-i\theta(t)}\;
\left[1+r(\tau)^2\,\gamma^{5}\,
\int_{-\infty}^\infty dv \sum_s h^{(s)}_{1}(v,\sigma)\right]
\nonumber\\
&&\label{eq:Ecompsg}\\
imkE_{mk}(t)&=&\gamma^{3m+2}\,r(\tau)^{m}\,e^{-im\theta(t)}\;\gamma^{2m-2}\,
\int_{-\infty}^\infty dv \sum_s h^{(s)}_{m}(v,\sigma)
\hspace{1.0in}m\geq2.
\nonumber
\end{eqnarray}
The asymptotic electric field is
\begin{equation}
\frac{E(x,t)}{\gamma^{5/2}}=\frac{1}{k}\left\{-{ir(\tau)e^{-i\theta(t)}}\;
\left[1+r(\tau)^2\,\gamma^{5}\,
\int_{-\infty}^\infty dv \sum_s h^{(s)}_{1}(v,\sigma)\right]\,e^{ikx}
+\mbox{\rm cc}\right\}
+{\cal O}({\gamma^{11/2}});
\end{equation}
clearly $E$ is dominated by the wavenumber of the unstable mode with
an overall scaling of $\gamma^{5/2}$. The term $\gamma^{5} \int dv
\sum_s h^{(s)}_{1}$ is treated as an ${\cal O}(1)$ contribution in
light of the estimate (\ref{eq:hmg}) above.

\subsubsection{Special cases: $c_1(0)=0$}

For instabilities with fixed ions or flat ion distributions, we have
$\beta=2$ and $\mu_{m,j}=J_{m,j}-j-\delta_{m,1}$ in (\ref{eq:hmj});
applying this bound to the series $h_m=\sum_j h_{m,j}\sigma^j$ yields
\begin{equation}
\lim_{\gamma\rightarrow0^+} \gamma^{\alpha_m} \left|\int^\infty_{-\infty}
dv \, \sum_s h^{(s)}_{m}(v)\right| < \infty
\label{eq:hmsc}
\end{equation}
with
\begin{equation}
\alpha_m=\left\{\begin{array}{lc}
1& m=0\\
3&m=1\\
2m-3&m\geq2
\end{array}\right.\hspace{0.5in}(c_1(0)=0).\label{eq:aspecial}
\end{equation}
Now the general expressions for the components reduce to
\begin{eqnarray}
ikE_k(t)&=&\gamma^{2}\,r(\tau)e^{-i\theta(t)}\;
\left[1+\gamma\,r(\tau)^2\,\gamma^{3}\,
\int_{-\infty}^\infty dv \sum_s h^{(s)}_{1}(v,\sigma)\right]
\nonumber\\
&&\label{eq:Ecompsc}\\
imkE_{mk}(t)&=&\gamma^{2m+3}\,r(\tau)^{m}\,e^{-im\theta(t)}\;\gamma^{2m-3}\,
\int_{-\infty}^\infty dv \sum_s h^{(s)}_{m}(v,\sigma)
\hspace{1.0in}m\geq2,
\nonumber
\end{eqnarray}
and the asymptotic electric field has the form
\begin{equation}
\frac{E(x,t)}{\gamma^{2}}=\frac{1}{k}\left\{-{ir(\tau)e^{-i\theta(t)}}\;
\left[1+  {\cal O}(\gamma)\right]\,e^{ikx}
+\mbox{\rm cc}\right\}
+{\cal O}({\gamma^{5}}).
\end{equation}
The overall scaling is now the well known $\gamma^2$ or ``trapping
scaling'' and the leading term has a much simpler structure. Again we
find the wavenumber $k$ of the linear instability; however now the
time dependence is simply given by the mode amplitude
$r(\tau)\exp(-i\theta(t))$.

\section{Discussion}\label{sec:disc}

The single wave model, derived originally by O'Neil, Winfrey and
Malmberg, described the interaction of a cold electron beam
interacting with a plasma of mobile electrons and fixed ions. In their
problem, the infinite extent of the plasma allowed for continuous
wavenumbers and the dispersion relation for a cold beam was required
to select a single wavenumber corresponding to the maximum growth
rate. This wavenumber characterizes the electric field whose nonlinear
time development results from the coupling to resonant particles. The
nonresonant plasma simply provides a linear dielectric which supports
the wave.

By contrast, we pose a more general problem, allowing for multiple
mobile species and not restricting the type of electrostatic
instability, but for a finite plasma with periodic boundary
conditions. Within this setting, we consider equilibria supporting a
single unstable mode and derive the resulting equations for the
electric field and distributions in the limit of weak instability. In
this asymptotic limit, the physical picture of the original single
wave model emerges quite generally. The monochromatic electric field
is coupled to the resonant particles and evolves nonlinearly while the
nonresonant particles show only a linear response to the electric
field.

The amplitude expansions, whose singularity structure form the basis
of our analysis, do not provide a practical tool for solving the
single wave model. For this purpose, it is more convenient to assume
the simplifications of the single wave picture and derive model
equations directly from the original Vlasov theory.  This development
will be presented in a forthcoming paper.\cite{jdcaj98}

\acknowledgments This work was supported by NSF grant PHY-9423583.

\end{document}